\documentclass[aps,preprint]{revtex4}%
\usepackage{amsfonts}
\usepackage{amsmath}
\usepackage{amssymb}
\usepackage{graphicx}%
\begin{document}
\title{Entropy Conservation of Linear Dilaton Black Holes in Quantum Corrected
Hawking Radiation}
\author{I.Sakalli}
\email{izzet.sakalli@emu.edu.tr}
\author{H.Pasaoglu}
\email{hale.pasaoglu@emu.edu.tr}
\author{M. Halilsoy}
\email{mustafa.halilsoy@emu.edu.tr}
\affiliation{Department of Physics, Eastern Mediterranean University,G. Magusa, North
Cyprus, Mersin-10, Turkey.}

\begin{abstract}
It has been shown recently that information is lost in the Hawking radiation
of the linear dilaton black holes in various \ theories when applying the
tunneling formulism without considering quantum gravity effects. In this
Letter, we recalculate the emission probability by taking into account of the
log-area correction to the Bekenstein-Hawking entropy and the statistical
correlation between quanta emitted. The crucial role of the black hole remnant
on the entropy conservation is highlighted. We model the remnant as a higher
dimensional linear dilaton vacuum in order to show that such a remnant model
cannot radiate and its temperature would be zero. In addition to this, the
entropy conservation in the higher dimensional linear dilaton black holes is
also discussed. In summary, we show in detail that the information can also
leak out from the linear dilaton black holes together with preserving
unitarity in quantum mechanics.

\end{abstract}
\maketitle

\section{Introduction}

Since the seminal works of Bekenstein \cite{Bekenstein} and Hawking
\cite{Hawking}, it is known that a black hole (BH) behaves as a thermodynamic
system, radiating as a black body with characteristic temperature and entropy.
Today, the thermodynamic properties of BHs are well understood such that there
are numerous independent derivations that all give the same quantitative
results: Hawking temperature $T_{H}=\frac{\kappa}{2\pi}$\ and
Bekenstein-Hawking entropy $S_{BH}=\frac{A_{h}}{4}$, where $\kappa$ is the
surface gravity and $A_{h}$ is the horizon area of the BH. However, in today's
physics Hawking's original derivation is considered only a skillful
application of quantum field theory in curved spacetime. Although it is not
directly connected to quantum gravity, nevertheless it has a strong impact on
quantum-gravitational problems toward understanding the statistical origin of
the BH entropy, and its natural consequence -- information concept
\cite{Shannon} in the BHs. With the traditional picture of the Hawking
radiation, which generates entangled pairs and the state of the outgoing
quanta is a mixed (pure thermal radiation) when the BH completely evaporates,
one encounters with two serious and simultaneous problems. One of them is the
violation of the unitarity in quantum mechanics (QM), which does not allow any
evolution process from a pure quantum state into a mixed state. The second is
the information loss about the original quantum state that formed the black
hole. The latter phenomenon is called "information loss paradox" \cite{ILP}.
In order to resolve this paradox, one finds many attempts in the literature,
see for instance \cite{RESILP1,ChenShao} (and references therein). Among them,
tunneling of emitted particles through the horizon is one of the popular
methods. The speech-ripe idea of the method dates back to 1999
\cite{Srinivasan}, but the nicest form was developed a decade ago by Parikh
and Wilczek \cite{ParikhWilczek}. In a very short time, Parikh and Wilczek's
tunneling method has been used and improved further by other researchers
\cite{Formulism}, and it has been known as the tunneling formulism. This
formulism has recently been improved one level up by Chen and Shao
\cite{ChenShao} by considering the statistical correlations, which exist among
the emitted quanta \cite{Zhang}. In \cite{ChenShao}, it has successfully been
shown that emitted particles would leak out information from the BH. Results
of \cite{ChenShao} do not only resolve the information loss paradox, but
preserve total entropy conservation and the unitarity of QM as well. All those
impressive results in \cite{ChenShao}\ have become possible with the inclusion
of the quantum gravity corrections and the existence of BH remnant.

Study \cite{ChenShao} has given us a hope also to resolve the information loss
problem in our recent study \cite{PaSa}, which has explored the Hawking
radiation of linear dilaton black holes (LDBHs) within the context of Maxwell,
Yang-Mills and Born-Infeld theories. In \cite{PaSa}, we have not considered
the BH remnant with quantum corrections. Our calculations yield a pure thermal
spectrum, which corresponds to the violation of the unitarity in QM and
signals the information loss in the Hawking radiation of the LDBHs.

In this Letter, following \cite{ChenShao}\ as a guide, we calculate the
correlation between successively emitted quanta from the LDBHs. The existence
of the LDBH remnant in preserving entropy conservation will be much more
important than the Schwarzschild case \cite{ChenShao}. In Section 2, we apply
the tunneling formulism to the LDBHs in order to obtain the quantum gravity
corrected BH entropy and the emission probability. Section 3 is devoted to the
calculation of the correlation between two successively emitted quanta. In the
following Section, complete radiation process is considered in which particles
are successively emitted from the LDBH until reaching the remnant with an
emphasis on the entropy conservation. Similarly, in Section 5, we consider the
case of higher dimensional LDBHs \cite{MSH}. First, we model the remnant as a
higher dimensional linear dilaton vacuum and show, from the absence of
reflection, that its temperature must be zero. Secondly, we reconsider the
generic higher dimensional LDBHs and calculate the entropy of the pointlike
remnant BH. We draw our conclusions in Section 6.

Throughout the Letter, the units $G=c=k_{B}=\hbar=1$ and the Planck length
$L_{p}=\sqrt{\frac{\hbar G}{c^{3}}}=1$ are used.

\section{Tunneling rate with quantum corrections for $4D$-LDBHs}

As it can be seen from \cite{PaSa}, $4D$-LDBHs in Einstein-Maxwell-Dilaton
(EMD), Einstein-Yang-Mills-Dilaton (EYMD) and
Einstein-Yang-Mills-Born-Infeld-Dilaton (EYMBID) theories are described by the
line element.%

\begin{equation}
ds^{2}=-fdt_{L}^{2}+\frac{dr^{2}}{f}+R^{2}d\Omega_{2}^{2},
\end{equation}

where $t_{L}$ is the LDBH time and $d\Omega_{2}^{2}=d\theta^{2}+\sin^{2}\theta
d\phi^{2}.$ Here, the metric functions are given by%

\begin{equation}
f=\Sigma r(1-\frac{r_{+}}{r})\text{ \ \ and \ \ \ }R=A\sqrt{r},
\end{equation}

It is obvious that metric (1) represents a static non-rotating BH with a
horizon at $r_{+}.$ The coefficients $\Sigma$ and $A$ in the metric functions
take different values according to the concerned theory (EMD, EYMD or EYMBID)
\cite{PaSa}. Furthermore, it is a non-asymptotically flat (NAF) spacetime. For
$r_{+}\neq0$, the horizon hides the null singularity at $r=0$. However, in the
extreme case $r_{+}=0,$ the central null singularity $r=0$ is marginally
trapped in which it does not allow outgoing signals to reach external
observers. Namely, even in the extreme case of $r_{+}=0$ metric (1) maintains
its black hole property.

By using the definition of quasi-local mass $M$ \cite{BrownYork} for our NAF
metric (1), one can get a relationship between the horizon $r_{+}$\ and the
mass $M$ as follows%
\begin{equation}
r_{+}=\frac{4M}{\Sigma A^{2}},
\end{equation}

Since none of the curvature invariants of metric (1) are singular on the
horizon $r_{+}$, one would transform to a new coordinate system, which can be
a non-singular at $r_{+}$. For this purpose, we pass to
Painlev\'{e}-Gullstrand type coordinates%

\begin{equation}
ds^{2}=-f(r)dt^{2}+2\sqrt{1-f(r)}dtdr+dr^{2}+R^{2}d\Omega^{2},
\end{equation}

with%

\begin{equation}
dt=dt_{L}+\frac{\sqrt{1-f(r)}}{f(r)}dr,
\end{equation}

where metric (4) has a number of advantages for such problems. Of course the
curvature singularity at the origin $r=0$ is still present in both coordinate
systems (1) and (4). Meanwhile, from the Schwarzschild problem it is known
that in these coordinates the time $t$ is linearly related to the proper time
for a radially falling observer. Obviously, the hypersurfaces $t=const.$ are
all intrisically flat.

In metric (4),\ the radial null geodesics of a test particle considered as a
massless spherical shell have a rather simple form,%

\begin{equation}
\dot{r}=\frac{dr}{dt}=-\sqrt{1-f(r)}\pm1,
\end{equation}

where the choice of signs depends on whether the rays are outgoing ($+$) or
ingoing ($-$). In Painlev\'{e}-Gullstrand coordinates (4), the surface gravity
on the horizon, which is the function of the BH mass $M$ is one of the
Christoffel components.%

\begin{equation}
\kappa(M)=\Gamma_{00}^{0}=\frac{1}{2}f^{\prime}(r_{+}),
\end{equation}

Since the metric function $f$ is zero on the horizon, it can be expanded
around $r_{+}$ as%

\begin{equation}
f=f^{\prime}(r_{+})(r-r_{+})+O(r-r_{+})^{2},
\end{equation}

By virtue of the $4D$-LDBHs the $O(r-r_{+})^{2}$\ vanishes, and we have an
exact expression.\ However, in the higher dimensional cases, which we consider
in Section V, an expression around $r_{+}$ will be used. Therefore, near the
horizon of the $4D$-LDBH, the radial outgoing null geodesics can be expressed as%

\begin{equation}
\dot{r}=\frac{dr}{dt}=\frac{1}{2}f^{\prime}(r_{+})(r-r_{+}),
\end{equation}

which is nothing but%

\begin{equation}
\dot{r}=(r-r_{+})\kappa(M),
\end{equation}

Let us consider a spherically symmetric system whose total mass $M$ is kept
fixed. We assume that the system consists of a LDBH with varying mass
$M-\omega,$ emitting a spherical shell of mass $\omega$ such that $\omega\ll
M$. This phenomena is known as self-gravitation effect \cite{Kraus}. After
taking into account of such self-gravitation effects, $\dot{r}$\ can be
rewritten as%

\begin{equation}
\dot{r}=(r-r_{+})\kappa(M-\omega),
\end{equation}

In the WKB approximation, the tunneling rate for an outgoing positive energy
particle, which crosses the horizon from $r_{in}$\ to $r_{out},$ is related to
the imaginary part of the particle's action \cite{Kraus,ParikhWilczek} in
accordance with%

\begin{equation}
\Gamma\sim e^{-2\operatorname{Im}(I)},
\end{equation}

Imaginary part of the particle's action is calculated from%

\begin{equation}
\operatorname{Im}(I)=\operatorname{Im}\int_{r_{in}}^{r_{out}}p_{r}%
dr=\operatorname{Im}\int_{r_{in}}^{r_{out}}\int_{0}^{p_{r}}d\tilde{p}_{r}dr,
\end{equation}

Using Hamilton's equation for the classical trajectory in the form%

\begin{equation}
dp_{r}=\frac{dH}{\dot{r}},
\end{equation}

with $H=M-\omega$\ \textit{i.e.} $dH=-d\omega$, and inserting $\dot{r}$ given
by Eq. (10) into Eq. (13), one gets%

\begin{equation}
\operatorname{Im}(I)=-\operatorname{Im}\int_{r_{in}}^{r_{out}}\int_{0}%
^{\omega}\frac{d\tilde{\omega}}{(r-r_{+})\kappa(M-\tilde{\omega})}dr,
\end{equation}

One can evaluate the integral over $r$ by deforming a contour, where its
semicircle centered at real axis pole $r_{+}.$ Thus we get%

\begin{equation}
\operatorname{Im}(I)=-\pi\int_{0}^{\omega}\frac{d\tilde{\omega}}%
{\kappa(M-\tilde{\omega})},
\end{equation}

The reason of the sign change in (15) is due to the fact that the horizon will
shrink during the process of Hawking Radiation, $r_{out}<r_{in}$. In other
words, the horizon tunnels inward as the BH's mass decreases.

According to corrections of surface gravity, the quantum gravity surface
gravity \cite{Zhang,Banerjee} is defined as%

\begin{equation}
\kappa_{QG}=\kappa(M-\omega),
\end{equation}

and Eq. (16) becomes%

\begin{equation}
\operatorname{Im}(I)=-\pi\int_{0}^{\omega}\frac{d\tilde{\omega}}{\kappa_{QG}},
\end{equation}

Hawking temperature is still expressed in the form $T_{H}=\frac{\kappa_{QG}%
}{2\pi}$. By using the Hawking temperature in Eq. (18), we find

\begin{align}
\operatorname{Im}(I)  &  =-\frac{1}{2}\int_{0}^{\omega}\frac{d\tilde{\omega}%
}{T_{H}}=-\frac{1}{2}\int_{S_{in}}^{S_{out}}dS\nonumber\\
&  =-\frac{1}{2}\int_{S_{QG}(M)}^{S_{QG}(M-\omega)}dS\nonumber\\
&  =\frac{1}{2}\left[  S_{QG}(M)-S_{QG}(M-\omega)\right] \\
&  =\frac{1}{2}\Delta S_{QG}.
\end{align}

where $S_{QG}$ is the corrected area entropy for BH. In string theory and loop
quantum gravity \cite{SQG} it is introduced with a logarithmic correction%

\begin{equation}
S_{QG}=\frac{A_{h}}{4}+\alpha\ln A_{h}+O(\frac{1}{A_{h}}),
\end{equation}

where $\alpha$ is a dimensionless constant. It takes different values
according to which theory is considered. The other physical parameter $A_{h}$
is the area of horizon $A_{h}$. For the $4D$-LDBHs, $A_{h}=4\pi A^{2}%
r_{+}=16\pi\frac{M}{\Sigma},$ and one easily reads the tunneling rate with
quantum correction as%

\begin{equation}
\Gamma\sim e^{-2\operatorname{Im}(I)}=e^{-\Delta S_{QG}}=(1-\frac{\omega}%
{M})^{\alpha}\exp\left(  -4\pi\frac{\omega}{\Sigma}\right)  .
\end{equation}
$\bigskip$

Here, $(1-\frac{\omega}{M})^{\alpha}$ is an additional factor compared with
the previous tunneling rate in \cite{PaSa}. This expression comes from quantum
gravitational effects on the energy of emitted particles and the mass of the
LDBH. In other words, it is natively due to the effect of back reaction. As it
will be seen in the following sections, it will play a great role on entropy
conservation as well as on the information loss paradox. One may notice that
the coefficient of power $\alpha$\ herein is 1, as distinct from 2 appeared in
the tunneling calculations of Schwarschild black hole, see for instance
\cite{ChenShao}.

\section{Statistical correlation between two successive emissions}

According to quantum gravity effects and its consequence (21), contrary to the
classical tunneling method employed in \cite{PaSa},\ we cannot consider
anymore the LDBHs as objects emitting only pure thermal radiation. From
physical point of view, the deviation (21) from thermal spectrum is very
important, and it may cause some therapeutic effects on the information loss
appeared in the LDBHs \cite{PaSa}. In this section, we shall check whether or
not the emission probabilities for two successive modes are statistically
correlated \cite{QCbook}. Similar to the conclusion of non-trivial correlation
obtained in \cite{ChenShao}, we expect to find statistically dependent
correlations between two successive emissions. By that way, we would like to
open a gate in order to resolve the information paradox in the LDBHs.
Throughout this section, we shall follow the method employed in
\cite{ChenShao}.

Firstly, we consider initially two successive emissions, with energies
$\omega_{1}$ and $\omega_{2}$. Using Eq. (21), for the first emission of
energy $\omega_{1}$ we have%

\begin{equation}
\Gamma(\omega_{1})=(1-\frac{\omega_{1}}{M})^{\alpha}\exp\left(  -4\pi
\frac{\omega_{1}}{\Sigma}\right)  ,
\end{equation}

Then a second emission of energy $\omega_{2}$\ on the condition that the first
one $\omega_{1}$ is%

\begin{equation}
\Gamma(\left.  \omega_{2}\right\vert \omega_{1})=(1-\frac{\omega_{2}}%
{M-\omega_{1}})^{\alpha}\exp\left(  -4\pi\frac{\omega_{2}}{\Sigma}\right)  ,
\end{equation}

which is known as the conditional probability \cite{Zhang}. Removing the
condition on the second emission, we get%

\begin{equation}
\Gamma(\omega_{2})=(1-\frac{\omega_{2}}{M})^{\alpha}\exp\left(  -4\pi
\frac{\omega_{2}}{\Sigma}\right)  ,
\end{equation}

\bigskip which is the probability just for the second emission. The emission
of the total energy yields%

\begin{equation}
\Gamma(\omega_{1}+\omega_{2})=(1-\frac{\omega_{1}+\omega_{2}}{M})^{\alpha}%
\exp\left(  -4\pi\frac{\omega_{1}+\omega_{2}}{\Sigma}\right)  ,
\end{equation}

The statistical correlation between two successive emissions is measured by
\cite{Parikh0402166}%

\begin{equation}
\chi(\omega_{1}+\omega_{2};\omega_{1},\omega_{2})=\ln\frac{\Gamma\left(
\omega_{1}+\omega_{2}\right)  }{\Gamma\left(  \omega_{1}\right)  \Gamma\left(
\omega_{2}\right)  },
\end{equation}

which is calculated as%

\begin{equation}
\chi(\omega_{1}+\omega_{2};\omega_{1},\omega_{2})=\alpha\ln(1-\frac{\omega
_{1}\omega_{2}}{\left(  M-\omega_{1}\right)  \left(  M-\omega_{2}\right)  }).
\end{equation}

First of all, it is evident that the correlation strictly depends on $\alpha$
. In the Schwarschild BH \cite{ChenShao}, even in the case of $\alpha=0,$\ the
correlation is non-zero. But, here once $\alpha=0$ is set to zero the
subsequent emissions become statistically independent, and thus information
paradox might never be resolved. Essentially, in the absence of $\alpha$ the
result (27) explains why we have not obtained any deviation from pure thermal
spectrum in our recent study \cite{PaSa}. However, the case $\alpha\neq0$ show
that the successive emissions are statistically dependent, which means that
there must be correlations between the successive emissions. In summary, Eq.
(27) indicates that information should leak out from the LDBHs during their radiation.

\section{Entropy conservation and BH remnant}

It is known that calculation of entropy carried by Hawking radiation is
obtained best by considering the complete process of the BH evaporation. For
this purpose, we consider the emission of $n$ particles with energies
$\omega_{1},\omega_{2},......,\omega_{n}$, which are successively emitted from
the LDBH. During the Hawking radiation, the BH loses its mass and
automatically its entropy, that is its total entropy part by part is
transferred to the emitted particles and their correlations.

As shown in \cite{ChenShao}, at the end of the evaporation we should have only
the BH remnant with energy $\omega_{c}$ such that $\omega_{1}+\omega
_{2},......+\omega_{n}+\omega_{c}=M$. By using the derived formula of the
total entropy carried out by radiation \cite{ChenShao}%

\begin{equation}
S=-\ln%
{\displaystyle\prod\limits_{i=1}^{n}}
\Gamma(M-\sum_{j=1}^{i-1}\omega_{j}\left\vert \omega_{i}\right.  ),
\end{equation}

one obtains%

\begin{align}
S  &  =-\ln\left\{  \left(  \frac{\omega_{c}}{M}\right)  ^{\alpha}\exp\left[
-\frac{4\pi}{\Sigma}(M-\omega_{c})\right]  \right\} \nonumber\\
&  =\frac{4\pi}{\Sigma}M+\alpha\ln(\frac{M}{\omega_{c}})-\frac{4\pi}{\Sigma
}\omega_{c},
\end{align}

Here, in order to avoid divergence of the entropy the non-zero $\omega_{c}$ is
of vital importance. So, we can infer that the quantum gravity corrected
entropy is not valid for mini black holes. In the literature, there are many
discussions about the remnant and its implication, see for instance
\cite{Remnant}. Constitutively, one can now see that it is a natural result of
the quantum gravity effects. Also the inclusion of quantum gravity effects is
in accordance with the generalized uncertainty principle, which might cease
the complete evaporation of the BH \cite{GUP}. Besides this, thinking of the
remnant as a non-radiate object having an infinitesimal surface area would not
be absurd. From this point of view, in the next section we shall consider the
remnant as an extreme LDBH with pointlike horizon and almost zero mass. It
will be shown that such a black hole cannot radiate and its temperature would vanish.

Furthermore, the entropy of the remnant can be calculated by applying the
Bekenstein's entropy bound \cite{BekensteinBound}. To find it, we reconsider
the entropy $S$\ carried out by radiation and then follow the following procedure%

\begin{align}
S  &  =\frac{4\pi}{\Sigma}M+\alpha\ln(\frac{\frac{16\pi M}{\Sigma}}%
{\frac{16\pi\omega_{c}}{\Sigma}})-\frac{4\pi}{\Sigma}\omega_{c}\nonumber\\
&  =\frac{4\pi}{\Sigma}M+\alpha\ln(\frac{16\pi M}{\Sigma})-\left[  \alpha
\ln(\frac{16\pi\omega_{c}}{\Sigma})+\frac{4\pi}{\Sigma}\omega_{c}\right]
\nonumber\\
&  =(\frac{A_{h}}{4}+\alpha\ln A_{h})-\left[  \alpha\ln(\frac{16\pi\omega_{c}%
}{\Sigma})+\frac{4\pi}{\Sigma}\omega_{c}\right] \nonumber\\
&  =S_{QG}-S_{C},
\end{align}

One can easily read the remnant's entropy as%

\begin{equation}
S_{C}=\alpha\ln(\frac{16\pi\omega_{c}}{\Sigma})+\frac{4\pi}{\Sigma}\omega_{c}.
\end{equation}

From Eq. (30), one can also deduce the conservation of entropy. Clearly, the
total entropy of a radiating LDBH $S_{QG}$ is equal to the entropy of black
hole remnant $S_{C}$\ plus the entropy carried out by radiation $S$. Being in
conform with other recent studies \cite{ChenShao,Zhang}, our result shows that
the exact spectrum of the Hawking radiation as a tunneling process is not a
pure thermal spectrum. This impressive conclusion also implies that
information is not lost, and unitarity in QM is restored during the Hawking radiation.

\section{Quantum corrected entropy of the remnant in higher dimensional LDBHs}

The generic line element for higher dimensional ($N\geq4$) static spherically
symmetric LDBHs in various theories can be best seen in \cite{MSH}. In higher
dimensions, the metric function $f$ of the LDBHs and the spherical line
element are modified to%

\begin{equation}
f=\Sigma r\left[  1-(\frac{r_{+}}{r})^{\frac{N-2}{2}}\right]  ,\text{
\ \ \ }d\Omega_{N-2}^{2}=d\theta_{1}^{2}+\underset{i=2}{\overset{N-3}{%
{\textstyle\sum}
}}\underset{j=1}{\overset{i-1}{%
{\textstyle\prod}
}}\sin^{2}\theta_{j}\;d\theta_{i}^{2},
\end{equation}

where $0\leq\theta_{k}\leq\pi$ with $k=1..N-3$, and $0\leq\theta_{N-2}\leq
2\pi.$ The modified form of the physical constant $\Sigma$\ in higher
dimensions can also be seen in \cite{MSH}.

As we mentioned before, we would like to set forth the remnant as a higher
dimensional LDBH with a pointlike horizon. So, the remnant can be thought as a
spacetime with negligible mass. From this point forth, the metric of the
remnant can be approximated to a linear dilaton vacuum metric as%

\begin{equation}
ds^{2}=-\Sigma rdt^{2}+\frac{dr^{2}}{\Sigma r}+R^{2}d\Omega_{N-2}^{2},
\end{equation}

One can find the statistical Hawking temperature \cite{Wald} of this metric as
a finite temperature with $T_{H}=\frac{\left(  N-2\right)  }{8\pi}\Sigma.$ But
this result is not persuasive since the remnant is massless, clearly a
candidate for a non-radiating object, expecting therefore its temperature as
zero will be more realistic. To this end, we proceed with a more precise
evaluation of the temperature of the remnant from the study of wave scattering
in such a spacetime. Metric (33) can be transformed into%

\begin{equation}
ds^{2}=\rho^{2}\left(  -d\tau^{2}+dx^{2}+d\Omega_{N-2}^{2}\right)  ,
\end{equation}

by%

\begin{equation}
r=e^{\beta x},\text{ \ \ \ \ }t=\frac{\beta}{\Sigma}\tau,\text{ \ \ \ \ \ \ }%
\rho=\frac{\beta}{\sqrt{\Sigma}}e^{\frac{\beta}{2}x}.
\end{equation}

where constant $\beta=A\sqrt{\Sigma}$ (for example, in $4D$-EMD theory it is 1
\cite{MSH,Clement}). Therefore metric (34) is conformal to the product of
$M_{2}\times S_{N-2}$ of a two-dimensional Minkowski space-time with the
$(N-2)$-sphere. The massless Klein-Gordon equation%

\begin{equation}
\nabla^{2}\Phi=0,
\end{equation}

with $\Phi=\rho^{^{-\left(  \frac{N-2}{2}\right)  }}\Psi$ can be reduced to%

\begin{equation}
-\rho^{^{\left(  \frac{N+2}{2}\right)  }}\left\{  \partial_{\tau\tau}%
-\partial_{xx}+\left[  \frac{\beta\left(  N-2\right)  }{4}\right]  ^{2}%
-\nabla_{N-2}^{2}\right\}  \Psi=0,
\end{equation}

where $\nabla_{N-2}^{2}$ is the $(N-2)$-dimensional Laplace-Beltrami operator
with the eigenvalue $-l(l+N-3)$ \cite{Du}. The reduced Klein-Gordon equation
can be rewritten in spherical harmonics with orbital quantum number $l$ as%

\begin{equation}
\nabla_{2}^{2}\Psi_{l}+\mu^{2}\Psi_{l}=0,
\end{equation}

where the effective mass $\mu$ can be found as%

\begin{equation}
\mu=\left\{  \left[  \frac{\beta\left(  N-2\right)  }{4}\right]
^{2}+l(l+N-3)\right\}  ^{\frac{1}{2}},
\end{equation}

In Eq. (38), $\nabla_{2}^{2}$ is the d'Alembertian operator on $M_{2}$. When
viewed from wave propagation aspect, the remnant (34) reduces to the
propagation of eigenmodes of a free Klein-Gordon field in two dimensions with
effective mass $\mu$. Conclusively, the remnant cannot radiate and therefore
contrary to the calculated $T_{H}=\frac{\left(  N-2\right)  }{8\pi}\Sigma
$\ value, its Hawking temperature should be zero.

Finally, we want to find the quantum corrected entropy of the remnant of the
LDBHs in an arbitrary dimension. For this reason, by using the surface area of
the higher dimensional LDBHs \cite{MSH}%

\begin{equation}
A_{h}=\frac{16\pi^{\frac{N-1}{2}}}{(N-2)\Gamma(\frac{N-1}{2})}\frac{M}{\Sigma
},
\end{equation}

one can find the dimension dependent modified area entropy of the BHs as%

\begin{equation}
S_{NQG}=\frac{A_{h}}{4}+\alpha\ln\left[  \frac{16\pi^{\frac{N-1}{2}}%
}{(N-2)\Gamma(\frac{N-1}{2})}\frac{M}{\Sigma}\right]  ,
\end{equation}

and read the dimension dependent quantum corrected tunneling rate%

\begin{equation}
\Gamma_{N}\sim e^{-2\operatorname{Im}(I)}=e^{-\Delta S_{NQG}}=(1-\frac{\omega
}{M})^{\alpha}\exp\left[  -\frac{4\pi^{\frac{N-1}{2}}}{(N-2)\Gamma(\frac
{N-1}{2})}\frac{\omega}{\Sigma}\right]  ,
\end{equation}

We notice that, higher dimensions do not change the correlation between two
successive emissions. That is to say, the correlation remains as in 4D case,
see Eq. (27). If we proceed to extend the study to the emission of $n$
particles with energies $\omega_{1},\omega_{2},......,\omega_{n}$, which are
successively emitted from the higher dimensional LDBHs, a straightforward
calculation will lead us to obtain the dimension dependent entropy carried out
by radiation as follows%

\begin{equation}
S_{N}=\frac{4\pi^{\frac{N-1}{2}}}{(N-2)\Gamma(\frac{N-1}{2})}\frac{M}{\Sigma
}+\alpha\ln(\frac{M}{\omega_{c}})-\frac{4\pi^{\frac{N-1}{2}}}{(N-2)\Gamma
(\frac{N-1}{2})}\omega_{c},
\end{equation}

This can be rearranged in the form%

\begin{equation}
S_{N}=S_{NQG}-S_{NC},
\end{equation}

where the dimension dependent entropy of the remnant $S_{NC}$ is found to be
\begin{equation}
S_{NC}=\alpha\ln\left[  \frac{16\pi^{\frac{N-1}{2}}}{(N-2)\Gamma(\frac{N-1}%
{2})}\frac{\omega_{c}}{\Sigma}\right]  +\frac{4\pi^{\frac{N-1}{2}}%
}{(N-2)\Gamma(\frac{N-1}{2})}\omega_{c}.
\end{equation}

Eq. (44) is again the conservation of entropy in higher dimensional LDBHs. We
conclude that even in the higher dimensional LDBHs information is not lost and
the unitarity of QM remains intact.

\section{Summary}

Taking the tunneling formulism and quantum gravity corrected entropy into
consideration, the radiation spectra of both $4D$ and higher dimensional LDBHs
are derived. Unlike our previous study \cite{PaSa}, here the probability
contains an overall factor with power $\alpha$. As it is shown in this Letter,
the role of the factor would be crucial in resolving the information paradox
for the LDBHs. By using the tunneling rate with quantum correction, the
correlation between successively emitted particles from both $4D$ and higher
dimensional LDBHs is found to be statistically dependent. This nontrivial
result proves that information should leak out from the LDBHs during their
radiation. When we consider the complete process of the black hole
evaporation, the conservation of entropy is obtained for LDBHs in all
dimensions and the importance of the black hole remnant proves to be decisive.
Absence of the remnant causes entropy divergence, and it is obviously not
preferable. Also it is shown that whenever the remnant is modeled as a
spacetime with a pointlike horizon -- almost zero mass -- a higher dimensional
linear dilaton vacuum metric can be used to describe it. Using the massless
Klein-Gordon equation, it is shown that such a remnant cannot radiate, as
expected, and its Hawking temperature would be zero.

In conclusion, the radiation spectrum of the LDBHs deviates from thermal
radiation whenever the quantum gravity corrections are taken into
consideration. That is, our results are consistent with the unitarity, and
show that the information is not lost in the process of Hawking radiation of
the LDBHs. Finally, we hope to find charged higher dimensional LDBHs and
extent our analysis to those as well.

\end{document}